\documentclass[a4paper,11pt]{article}
\usepackage{jheppub}
\usepackage{amsmath,amssymb,bm,mathrsfs}
\usepackage{mathtools,mathptmx,array,slashed}
\usepackage{multirow}
\usepackage[colorlinks=true,linkcolor=blue,citecolor=blue,urlcolor=blue]{hyperref}

\DeclareSymbolFont{symbols}{OMS}{cmsy}{m}{n}

\def\bse{\begin{subequations}}
\def\ese{\end{subequations}}
\def\bea{\begin{eqnarray}}
\def\eea{\end{eqnarray}}
\def\be{\begin{equation}}
\def\ee{\end{equation}}
\def\nn{\nonumber}
\def\p{\partial}

\newcommand{\cV}{\mathcal{V}}
\newcommand{\cL}{\mathcal{L}}
\newcommand{\cN}{\mathcal{N}}

\begin{document}

\title{Four-charge static non-extremal black holes in the five-dimensional $\cN =2$,
$STU-W^2U$ supergravity}

\author{Di Wu}
\author{and Shuang-Qing Wu}

\affiliation{ School of Physics and Astronomy, China West Normal University, \\
1 Shida Road, Nanchong, Sichuan 637002, People's Republic of China}

\emailAdd{wdcwnu@163.com, sqwu@cwnu.edu.cn}

\abstract{We construct, for the first time, new static non-extremal five-dimensional black
hole solutions (without or with squashed horizons) endowing with four different electric charge
parameters in the $D = 5$, $\cN = 2$ supergravity coupled to \emph{three} vector multiplets with
a specific pre-potential $\cV = STU -W^2U \equiv 1$. When the fourth charge parameter disappears,
the solution simplify reduces to the three-charge static black hole solution previously presented
in ref. \cite{PLB383-151}, which belongs to the solution to the $D = 5$, $\cN = 2$ supergravity
coupled to \emph{two} vector multiplets (also notably known as the $STU$ model). We parameterize
the model in such a simple fashion that not only can one easily recover the static three-charge
solution but also it is very convenient to study their thermodynamical properties of the obtained
black hole solutions in the case without a squashing horizon. We then show that the thermodynamical
quantities perfectly obey both the differential first law and integral Smarr formula of
thermodynamics. Finally, we also extend to present its generalizations with squashed horizons
or including a nonzero cosmological constant.}


\arxivnumber{2510.13655}

\maketitle

\flushbottom

\section{Introduction}

Exact solutions that represent black holes play a prominent role in General Relativity. Constructing
exact black hole solutions and analyzing their properties provides valuable insight into the nature
of (super)gravity and the structure of spacetime. The distinctive and often exotic features of black
holes in higher dimensions and supergravity theories, which are absent in the four-dimensional
solutions, have attracted considerable attention in recent years.

In this paper, we focus on constructing new static non-extremal four-charge black hole solutions
in the five-dimensional $\cN = 2$ supergravity and studying their thermodynamic properties. The
bosonic sector of five-dimensional $\cN = 2$ supergravity theory includes $n$ Abelian vector
multiplets in addition to the graviton and graviphoton. The interactions among the vectors
are fully specified by a symmetric tensor $C_{IJK}$, where the indices $I, J, K$ run over all
$(n+1)$ vector fields, including the graviphoton. Supersymmetry imposes strong constraints
on the admissible scalar manifolds, restricting them to nonlinear sigma models based upon
the so-called ``very special geometry" \cite{PLB293-94}, which plays a crucial role in the
classification of consistent supergravity theories in five dimensions.

The most studied solutions in the $D = 5$, $\cN = 2$ supergravity theory arise in the $U(1)^3$
case with $n = 2$ Abelian vector multiplets, commonly known as the ``$STU$" model. This model,
which couples the gravity multiplet to \emph{two} Abelian vector multiplets, admits a range of
exact black hole and black ring solutions due to its hidden symmetries and solution-generating
techniques. The static three-charge black hole was first constructed in 1996 by Horowitz,
Maldacena, and Strominger (HMS) \cite{PLB383-151}, and was then extended to the rotating case
by Cveti\v{c} and Youm \cite{NPB476-118,PRD54-2612}. Subsequently, the static solution was
generalized in ref. \cite{NPB553-317} to asymptotically AdS case with a nonzero negative
cosmological constant. Further extensions to construct the general non-extremal rotating charged
AdS$_5$ black hole solutions in the five-dimensional $U(1)^3$ gauged supergravity theory proved
challenging. Previously constructed non-extremal rotating charged AdS$_5$ black hole solutions
are limited to the two special cases: either with equal rotation parameters, or with some charges
equal. In the formal simpler situation where two rotation parameters are set equal, the solution
with three independent charges was obtained in ref. \cite{PRD70-081502}, extending
that was previously found in the minimal gauged supergravity \cite{PLB598-273}. For black holes
with two independent rotation parameters, a non-extremal solution where two charges are equal
but the third one is set to zero was found in ref. \cite{PRD72-041901}, and was then extended
\cite{PLB658-64} to the case in which two of the three charges are set equal, with the third
non-vanishing. In addition, a solution with only one charge non-zero was constructed
in ref. \cite{PLB644-192}, extending the special case given in ref. \cite{PRD72-041901} with only
one rotation parameter. The most general non-extremal solution (``Wu black hole" \cite{PLB707-286})
with three independent charges and two angular momenta was eventually obtained in 2012 by using an
extraordinarily useful ans\"atz that generalizes the usual Kerr-Schild one to string/supergravity
theory. Recent explorations into the $STU$ model have investigated its squashing versions
\cite{1009.3568,PLB726-404}, thermodynamic properties \cite{PLB383-151,NPB476-118,PRD54-2612,
NPB553-317,PRD70-081502,PRD72-041901,PLB658-64,PLB707-286}, hidden symmetries \cite{1608.05052},
and related structures \cite{JHEP0912001,JHEP1018080,EPJC77-96}, etc.

On the other hand, compared with the above-mentioned non-extremal solutions, due
to the fact that a set of efficient algorithms was developed to derive the BPS and almost-BPS
solutions, there exist a lot of extensive researches along the line of the BPS black hole solutions
to the five-dimensional minimal and $U(1)^3$ ungauged and gauged super-gravity theories, see for
example, refs. \cite{PLB391-93,CQG16-1,PLB503-147,CQG20-4587,PRD68-105009,JHEP0204006,MPLA13-239,
PLB442-97,PLB498-123,JHEP0201031,JHEP0404048,JHEP1005039,JHEP0107020,JHEP1007093,hep-th/0504080}.
Besides these mature studies within the $STU$ model, it is remarkable that a general three-charge
supersymmetric doubly-rotating AdS$_5$ black holes in the gauged $U(1)^3$ supergravity has been
found in ref. \cite{JHEP0406036}, based up the BPS case of the famous CCLP solution \cite{PRL95-161301}
in the five-dimensional minimal gauged Einstein-Maxwell-Chern-Simons supergravity.

Although considerable progress has been made in constructing new black hole solutions in the
five-dimensional $\cN = 2$ supergravity over the past several years, most of these results remain
confined to the $STU$ model, characterized by three independent charges.\footnote{
There are some attempts to include three extra (complex) scalar fields to explore the bubble AdS$_5$
or hairy extensions of the known solutions within the STU model \cite{PLB614-96,JHEP1111035}. Along
this direction, some recent progress has been made to find black hole solutions of five-dimensional
$U(1)^3$ gauged supergravity with three additional complex scalar fields $(\Phi_{1,2,3} \neq 0)$, see,
for instance, \cite{CQG36-02LT01,JHEP0319110,JHEP0523053,JHEP0625051}.} In contrast, less is known
when the number of vector multiplets exceeds two. It is plausible that further classes of black hole
solutions remain undiscovered in the general setting of the five-dimensional $\cN = 2$ supergravity.
In 2012, Giusto and Russo \cite{CQG29-085006} introduced a fourth charge to supersymmetric black ring
solutions by treating it perturbatively \cite{SPP144-255} and subsequently uplifted these configurations
to the eleven-dimensional supergravity via various string dualities. The resulting geometry can be
consistently truncated to the five-dimensional $\cN = 2$ supergravity coupled to \emph{three} vector
multiplets. This setup extends the $STU$ model by including one additional Abelian vector field and
its associated scalar, and is therefore structurally more intricate. At present, to the best of the
authors' knowledge, no known solution-generating techniques appear applicable in this broader context.
Several works \cite{CQG29-085006,JHEP0512033,JHEP0712107} have addressed supersymmetric black rings
with four electric and four dipole charges in this extended framework. However, no result has been
reported on the non-extremal black hole or black ring solutions beyond the supersymmetric limit.

In this article, we shall consider the five-dimensional $\cN = 2$ supergravity theory coupled
to \emph{three} Abelian vector multiplets ($n=3$), which we will dub as the $STU-W^2U$ model
according to its pre-potential. We then present new static, non-extremal black hole solutions
that carry four independent electric charges, thereby extending the known solution-space of the
five-dimensional supergravity beyond the conventional $STU$ framework. The remaining part of
this paper is organized as follows. Section 2 briefly reviews the five-dimensional $\cN = 2$
supergravity, focusing on the well-known $STU$ model and its associated black hole solutions.
In Sec. 3, we introduce the $STU-W^2U$ model and present its Lagrangian and equations of motion
by using a particularly useful parametrization of the scalar manifold. After that, we then
construct various static black hole solutions carrying four electric charges. For the general
case with four different electric charges, we compute its conserved mass and discuss its
thermodynamic property. We then show that together with the entropy, Hawking temperature,
the four electric charges and their corresponding electrostatic potentials, these quantities
completely satisfy both the differential and integral forms of the first law of black hole
thermodynamics. Section 4 summarizes our findings and outlines possible directions for future
work.

\section{$D = 5$, $\cN = 2$ ungauged supergravity and $STU$ model}

In this section, we provide a concise overview of the five-dimensional $\cN = 2$ supergravity, with
particular emphasis on the ungauged $STU$ model and its static black hole solutions. This framework
serves as the foundation for our subsequent construction of the four-charge generalization.

\subsection{Basic framework}

The bosonic sector of $D = 5$, $\cN = 2$ ungauged supergravity coupled to $n$ vector multiplets
is governed by the Lagrangian \cite{NPB242-244,NPB253-573}:
\be
\widehat{\cL} = \sqrt{-g}\Big[R -\frac{1}{2} Q_{IJ} F^I_{\mu\nu}F^{J\mu\nu}
 -Q_{IJ}(\p_{\mu} X^I)\p^{\mu} X^J\Big] -\frac{1}{24}C_{IJK}F^I_{\mu\nu}
 F^J_{\rho\sigma} A_{\lambda}^K\epsilon^{\mu\nu\rho\sigma\lambda} \, ,
\label{eq:lagrangian}
\ee
where $I, J = 1, \dots,\, n+1$, $R$ denotes the Ricci scalar curvature, $F_{\mu\nu}^I$ represent
the Abelian field strength tensors, and $X^I$ parameterize the scalar manifold. The constant
symmetric tensor $C_{IJK}$ plays a crucial role in ensuring gauge invariance of the Chern-Simons
term. The theory encompasses $n+1$ vector fields in total, comprising $n$ Abelian gauge fields
from the vector multiplets and the graviphoton from the supergravity multiplet.

The seminal work of refs. \cite{NPB242-244,NPB253-573} established the general framework for these
theories through an ans\"atz that depends generically on scalar fields. By demanding supersymmetry
invariance and closure of the supersymmetry algebra, they derived a set of algebraic and differential
constraints. The most general solution to these constraints introduces an auxiliary ambient space
with coordinates $X^I$ and defines a cubic pre-potential:
\be
\cV = \frac{1}{6}C_{IJK}X^IX^JX^K \equiv 1 \, .
\label{pre-potential}
\ee
This pre-potential induces a symmetric metric on the ambient space:
\be
Q_{IJ} = -\frac{1}{2} \p_I\p_J \ln\cV \big|_{\cV=1} \, ,
\ee
where $\p_I$ denotes the partial differentiation with respect to the $X^I$ associated with the
physical scalar fields $\varphi^i$. Remarkably, all quantities in the Lagrangian \eqref{eq:lagrangian}
can be expressed in terms of this pre-potential, defining what is known as ``very special geometry"
\cite{PLB293-94}.

\subsection{The $STU$ model}

Among the various possibilities within this framework, the $STU$ model stands out as a particularly
important and well-studied example. This model corresponds to $\cN = 2$ supergravity coupled to
\emph{two} Abelian vector multiplets, characterized by the pre-potential:
\be
\cV = X^1X^2X^3 = STU \equiv 1\, ,
\ee
where we have identified $\{X^1, X^2, X^3\} = \{S, T, U\}$. The ambient space metric and its inverse
take particularly simple diagonal forms:
\be
(Q_{IJ}) = \text{diag} \Big(\frac{1}{2S^2}, \frac{1}{2T^2}, \frac{1}{2U^2}\Big) \, , \quad
(Q^{IJ}) = \text{diag} \big(2S^2, 2T^2, 2U^2\big) \, .
\ee

To connect with more familiar field theory expressions, we parameterize the scalars in terms of two
dilaton scalar fields $(\varphi_1, \varphi_2)$:
\be
S = e^{\varphi_1 +\varphi_2}\, , \quad T = e^{\varphi_1 -\varphi_2}\, , \quad U = e^{-2\varphi_1} \, .
\ee
In terms of this parametrization, the Lagrangian for the $STU$ model assumes the familiar form:
\be\begin{split}
L &= \sqrt{-g}\Big[ R -3(\p_\mu\varphi_1)\p^\mu\varphi_1 -(\p_\mu\varphi_2)\p^\mu\varphi_2
 -\frac{1}{4} e^{-2\varphi_1 -2\varphi_2}F^1_{\mu\nu} F^{1\mu\nu} \\
&\quad -\frac{1}{4}e^{-2\varphi_1 +2\varphi_2}F^2_{\mu\nu}F^{2\mu\nu}
 -\frac{1}{4}e^{4\varphi_1}F^3_{\mu\nu} F^{3\mu\nu} \Big]
 -\frac{1}{4} \epsilon^{\mu\nu\rho\sigma\lambda}F^1_{\mu\nu}F^2_{\rho\sigma}A^3_\lambda \, ,
\end{split}\ee
where $F^I = dA^I\equiv F^I_{\mu\nu}dx^{\mu}\wedge dx^{\nu}$ are the field strength 2-forms of the
three $U(1)$ gauge fields. The Chern-Simons term, while not contributing in static configurations,
is also included here for completeness as it plays an important role in the rotating solutions to
ensure the consistency of the supersymmetric theory.

\subsection{Static non-extremal $STU$ black hole}

The $STU$ model admits a rich family of black hole solutions. Here we only mention two static,
non-extremal black holes with three independent electric charges, namely, the HMS solution
\cite{PLB383-151} and its squashed counterpart \cite{PLB726-404}.

The general static three-charge HMS black hole solution takes the form:
\be\begin{split}
ds^2 &= (H_1H_2H_3)^{1/3} \Big( -\frac{1 -2m/r^2}{H_1H_2H_3} dt^2
 +\frac{dr^2}{1 -2m/r^2} +r^2d\Omega_3^2\Big) \, , \\
A_I &= \frac{2mc_Is_I}{r^2H_I} dt \, , \quad  X^I = \frac{(H_1 H_2 H_3)^{1/3}}{H_I} \, ,
\end{split}\ee
where the harmonic functions $H_I = 1 +2ms_I^2/r^2$ encode the dependence on the charge parameters
$\delta_I$ with the constraint $c_I^2 =1 +s_I^2$ where $c_I = \cosh(\delta_I)$ and $s_I =
\sinh(\delta_I)$. The metric on the unit 3-sphere is given by:
\be
d\Omega_3^2 = d\theta^2 +\sin^2\theta\, d\phi^2 +\cos^2\theta\, d\psi^2 \, .
\label{3dsphere}
\ee

This solution represents a non-extremal black hole carrying three independent electric charges.
When all three charges vanish, it reduces to the five-dimensional Schwarzschild-Tangherlini solution,
while the extremal limit corresponds to $m \to 0$ with the charges held fixed. When three charge
parameters become identical, the solution recovers the five-dimensional Reissner-Nordstr\"om black
hole. On the other hand, directly applying the squashing transformation \cite{PTP116-417} to the
above HMS solution, one can get its squashed version. However, due to the non-vanishing of the
scalar moduli asymptotically at infinity, the first law generally acquires the contribution of
the scalar hairs \cite{PRL77-4992}. By contrast, via a brute-force method, one can also obtain
a relatively simple solution in which two scalar fields vanish at the asymptotical infinity,
as did in ref. \cite{PLB726-404}.

The structure of the above solution, with its characteristic product of harmonic functions,
has inspired numerous generalizations, such as the static extension to AdS$_5$ spacetime
\cite{NPB553-317}. The exact solutions in the ungauged $STU$ supergravity theory exhibit
the remarkable properties of very special geometry and can be systematically derived using
solution-generating techniques, making the $STU$ model an ideal testing ground for exploring
the interplay between black hole physics and supergravity.

\section{Four-charge static non-extremal black hole within the $STU-W^2U$ model}

Having reviewed the well-established $STU$ model, we now turn to our main objective: the construction
of static non-extremal black hole solutions with four independent electric charges. This requires
to extend the theoretical framework to incorporate an additional vector multiplet in the $STU-W^2U$
model.

\subsection{The $STU-W^2U$ model: Motivation and structure}

The quest for black hole solutions with more than three charges in the five-dimensional $\cN = 2$
supergravity has been a challenging endeavor. While the $STU$ model has been extensively studied,
models with additional vector multiplets offer the possibility of more general charge configurations.
Our approach builds upon insights from Giusto and Russo \cite{CQG29-085006}, who introduced a fourth
charge extension in the context of black ring solutions. However, we shall adjust their framework
for the $D = 5$, $\cN = 2$ supergravity coupled to $n=3$ Abelian vector multiplets and try to
construct various static non-extremal black holes.

The $STU-W^2U$ model is still defined by the pre-potential (\ref{pre-potential}) but now the
non-zero components of the symmetric tensor $C_{IJK}$ are being given by:
\be
C_{123} = 1\, , \quad C_{344} = C_{434} = C_{443} = -2 \, .
\ee
This choice represents a minimal extension of the $STU$ model in order to introduce a fourth
independent charge while maintaining the cubic structure of the pre-potential.

Identifying $\{X^1, X^2, X^3, X^4\} = \{S, T, U, W\}$, our pre-potential can be written as:
\be
\cV = X^1 X^2 X^3 -(X^4)^2 X^3 = STU -W^2 U \equiv 1.
\ee
This expression justifies our nomenclature ``$\mathit{STU-W^2U}$ model" and clearly shows that the
standard $STU$ model is recovered when $W = 0$. The ambient space metric derived from the above
pre-potential now is given as follows:
\be
(Q_{IJ}) =
\begin{pmatrix}
\frac{1}{2}T^2U^2 & \frac{1}{2}W^2U^2 & 0 & -TWU^2 \\
\frac{1}{2}W^2U^2 & \frac{1}{2}S^2U^2 & 0 & -SWU^2 \\
0 & 0 & \frac{1}{2U^2} & 0 \\
-TWU^2 & -SWU^2 & 0 & STU^2 +W^2U^2
\end{pmatrix} \, ,
\ee
with its inverse being given by:
\be
(Q^{IJ}) =
\begin{pmatrix}
2S^2 & 2W^2 & 0 & 2SW \\
2W^2 & 2T^2 & 0 & 2TW \\
0 & 0 & 2U^2 & 0 \\
2SW & 2TW & 0 & ST +W^2
\end{pmatrix} \, .
\ee
The non-diagonal nature of these matrices reflects the non-trivial mixing between the new field
$W$ and the original $STU$ sectors, which is a hallmark of this extended model.

\subsection{Lagrangian and field equations}

The construction of explicit solutions in the $STU-W^2U$ model requires a careful treatment of the
field equations derived from the Lagrangian. We begin by introducing a convenient parametrization
of the scalar fields that simplifies the subsequent analysis.

\paragraph{Scalar field parametrization:}
To facilitate the construction of explicit solutions, we parameterize the four scalars $X^I$ in
terms of three dilaton scalar fields $(\varphi_1, \varphi_2, \alpha)$ as follows:
\be
S = \sqrt{\alpha}e^{\varphi_1 +\varphi_2}\, , \quad T = \sqrt{\alpha} e^{\varphi_1 -\varphi_2}\, ,
\quad U = e^{-2\varphi_1} \, , \quad W = \sqrt{\alpha-1} e^{\varphi_1} \, .
\ee
This parametrization is chosen to diagonalize the kinetic terms as much as possible while maintaining
a clear connection to the $STU$ model limit (namely, $\alpha\to 1$). The scalar field $\alpha$ plays
a crucial role in incorporating the fourth scalar field $W$ while preserving the constraint
$\mathcal{V} \equiv 1$.

\paragraph{Bosonic Lagrangian:}
With the help of this strategy, the complete bosonic Lagrangian for $D = 5$, $\cN = 2$ ungauged
supergravity coupled to \emph{three} Abelian vector multiplets becomes:
\bea
\cL &=& \sqrt{-g}\bigg\{ R -3(\p_\mu\varphi_1)\p^\mu\varphi_1 -\alpha(\p_\mu\varphi_2)
 \p^\mu\varphi_2 -\frac{1}{4\alpha(\alpha-1)}(\p_\mu\alpha)\p^\mu\alpha
  -\frac{1}{4}e^{4\varphi_1}F^3_{\mu\nu}F^{3\mu\nu} \nn \\
&& -\frac{\alpha}{4} e^{-2\varphi_1}\big(e^{-2\varphi_2}F^1_{\mu\nu}F^{1\mu\nu}
 +e^{2\varphi_2}F^2_{\mu\nu}F^{2\mu\nu}\big) -\frac{1}{2}e^{-2\varphi_1}
 \Big[(2\alpha-1)F^4_{\mu\nu}F^{4\mu\nu} +(\alpha-1)F^1_{\mu\nu}F^{2\mu\nu}\Big] \nn \\
&&~~ +\sqrt{\alpha(\alpha-1)}e^{-2\varphi_1}\big(e^{-\varphi_2}F^1_{\mu\nu}
 +e^{\varphi_2}F^2_{\mu\nu}\big)F^{4\mu\nu} \bigg\} -\frac{1}{4}\epsilon^{\mu\nu\rho\sigma\lambda}
 \big(F^1_{\mu\nu}F^2_{\rho\sigma} -F^4_{\mu\nu}F^4_{\rho\sigma}\big)A^3_\lambda \, ,
\eea
where $F^I = dA^I\equiv F^I_{\mu\nu}dx^{\mu}\wedge dx^{\nu}$ are the field strength 2-forms of the
four $U(1)$ gauge field 1-forms $A^I = A^I_{\mu}dx^{\mu}$. The Lagrangian exhibits several noteworthy
features: the kinetic terms for the gauge fields show a non-trivial coupling to the scalar fields,
the four gauge fields interact through both minimal and non-minimal couplings, and the Chern-Simons
term now includes contributions involving the fourth gauge field $A^4$.

\paragraph{Dual field strengths:}
The modified 2-form fields, which play a key role in the equations of motion and charge definitions,
are given by:
\be\begin{split}
\widetilde{F}^1 &= e^{-2\varphi_1 -2\varphi_2}\alpha\, F^1 +e^{-2\varphi_1}(\alpha-1)F^2
 -2e^{-2\varphi_1 -\varphi_2}\sqrt{\alpha(\alpha-1)} F^4 \, , \\
\widetilde{F}^2 &= e^{-2\varphi_1}(\alpha-1)F^1 +e^{-2\varphi_1 +2\varphi_2}\alpha F^2
 -2e^{-2\varphi_1 +\varphi_2}\sqrt{\alpha(\alpha-1)}F^4 \, , \\
\widetilde{F}^3 &= e^{4\varphi_1}F^3 \, , \\
\widetilde{F}^4 &= \sqrt{\alpha(\alpha-1)} \big(e^{-2\varphi_1 -\varphi_2}F^1
 +e^{-2\varphi_1 +\varphi_2}F^2 \big) -e^{-2\varphi_1}(2\alpha-1)F^4 \, .
\end{split}\ee
Their dual 2-form fields satisfy the following generalized Bianich identities, which incorporate
the Chern-Simons couplings:
\be\begin{split}
d\big({}^\star \widetilde{F}^1\big) +F^2\wedge F^3 &= 0 \, , \qquad
d\big({}^\star \widetilde{F}^2\big) +F^1\wedge F^3 = 0 \, , \\
d\big({}^\star \widetilde{F}^3\big) +F^1\wedge F^2 -F^4\wedge F^4 &= 0 \, , \qquad
d\big({}^\star \widetilde{F}^4\big) +F^3\wedge F^4 = 0 \, ,
\end{split}\ee
in which a star represents the Hodge duality operation.

\paragraph{Gauge field equations:}
The four Abelian gauge field equations derived from the Lagrangian take the form:
\be
 \nabla_\nu F^{I\mu\nu}_{\text{cs}} \equiv
 \frac{1}{\sqrt{-g}}\p_\nu\big(\sqrt{-g}F^{I\mu\nu}_{\text{cs}}\big) = 0 \, ,
\ee
where the field strength tensors $F^I_{\text{cs}}$ incorporate both the modified field strengths
and Chern-Simons contributions:
\be\begin{split}
F^{1\mu\nu}_{\text{cs}} &= \widetilde{F}^{1\mu\nu} -\frac{1}{4}\varepsilon^{\mu\nu\rho\sigma\lambda}
 \big(F^2_{\rho\sigma}A^3_\lambda +F^3_{\rho\sigma}A^2_\lambda\big) \, , \\
F^{2\mu\nu}_{\text{cs}} &= \widetilde{F}^{2\mu\nu} -\frac{1}{4}\varepsilon^{\mu\nu\rho\sigma\lambda}
 \big(F^1_{\rho\sigma}A^3_\lambda +F^3_{\rho\sigma}A^4_\lambda\big) \, , \\
F^{3\mu\nu}_{\text{cs}} &= \widetilde{F}^{3\mu\nu} -\frac{1}{4}\varepsilon^{\mu\nu\rho\sigma\lambda}
 \big(F^1_{\rho\sigma}A^2_\lambda +F^2_{\rho\sigma}A^4_\lambda -2F^4_{\rho\sigma}A^4_\lambda\big) \, , \\
F^{4\mu\nu}_{\text{cs}} &= \widetilde{F}^{4\mu\nu} -\frac{1}{4}\varepsilon^{\mu\nu\rho\sigma\lambda}
 \big(F^3_{\rho\sigma}A^4_\lambda +F^4_{\rho\sigma}A^3_\lambda\big) \, .
\end{split}\ee
These equations demonstrate more intricate coupling between the four gauge fields in the
$STU-W^2U$ model, compared with the $STU$ theory.

\paragraph{Einstein equations:}
The contracted Einstein field equations, which govern the gravitational sector, are given by:
\bea
R_{\mu\nu} &=& 3(\p_\mu\varphi_1)\p_\nu\varphi_1 +\alpha(\p_\mu\varphi_2)\p_\nu\varphi_2
 +\frac{(\p_\mu\alpha)\p_\nu\alpha}{4\alpha(\alpha-1)}
 +\frac{\alpha}{2}e^{-2\varphi_1}\big(e^{-2\varphi_2}T^{11}_{\mu\nu}
 +e^{2\varphi_2}T^{22}_{\mu\nu}\big) +\frac{1}{2}e^{4\varphi_1}T^{33}_{\mu\nu} \nn \\
&& +e^{-2\varphi_1} \big[(\alpha-1)T^{12}_{\mu\nu} +(2\alpha-1)T^{44}_{\mu\nu}\big]
 -2\sqrt{\alpha(\alpha-1)} e^{-2\varphi_1} \big(e^{-\varphi_2} T^{14}_{\mu\nu}
 +e^{\varphi_2} T^{24}_{\mu\nu} \big) \, ,
\eea
where the contracted energy-momentum tensors are defined as:
\be
T^{IJ}_{\mu\nu} = \frac{1}{2}\big(F^I_{\mu\lambda}F^{J\lambda}_\nu
 +F^J_{\mu\lambda}F^{I\lambda}_\nu\big) -\frac{1}{6}g_{\mu\nu}F^I_{\rho\sigma}F^{J\rho\sigma} \, ,
\quad (I,J = 1, 2, 3, 4) \, .
\ee
The right-hand side of the Einstein equations clearly shows how all four gauge fields and three
scalar fields contribute to the stress-energy tensor that sources the curvature.

\paragraph{Scalar field equations:}
Finally, the equations of motion for three scalar fields $(\varphi_1, \varphi_2, \alpha)$ constitute
the system:
\be\begin{split}
&\hspace*{-1.2cm}
\frac{1}{\sqrt{-g}}\p_\mu\big(\sqrt{-g}\p^\mu\varphi_1\big) +\frac{\alpha}{12}e^{-2\varphi_1}
 \big(e^{-2\varphi_2}F^1_{\mu\nu}F^{1\mu\nu} +e^{2\varphi_2}F^2_{\mu\nu}F^{2\mu\nu}\big) \\
&\hspace*{-0.5cm} -\frac{1}{6}e^{4\varphi_1}F^3_{\mu\nu}F^{3\mu\nu}
 -\frac{1}{3}\sqrt{\alpha(\alpha-1)} e^{-2\varphi_1}\big(e^{-\varphi_2}F^1_{\mu\nu}
  +e^{\varphi_2}F^2_{\mu\nu}\big)F^{4\mu\nu} \qquad\quad \\
& +\frac{1}{6}e^{-2\varphi_1}\big[(\alpha-1)F^1_{\mu\nu}F^{2\mu\nu}
 +(2\alpha-1)F^4_{\mu\nu}F^{4\mu\nu}\big] = 0 \, ,
\end{split}\ee
\vskip -0.5cm
\be\begin{split}
&\frac{1}{\sqrt{-g}}\p_\mu \big(\alpha\sqrt{-g}\p^\mu\varphi_2\big)
 +\frac{\alpha}{4}e^{-2\varphi_1}\big(e^{-2\varphi_2}F^1_{\mu\nu}F^{1\mu\nu}
 -e^{2\varphi_2}F^2_{\mu\nu}F^{2\mu\nu}\big) \qquad\qquad \\
&\quad -\frac{1}{2}\sqrt{\alpha(\alpha-1)}e^{-2\varphi_1}\big(e^{-\varphi_2}F^1_{\mu\nu}
 -e^{\varphi_2}F^2_{\mu\nu}\big)F^{4\mu\nu} = 0 \, ,
\end{split}\ee
\vskip -0.5cm
\be\begin{split}
&\frac{1}{\sqrt{\alpha(\alpha-1)}\sqrt{-g}}\p_\mu\bigg[\frac{\sqrt{-g}}{\sqrt{\alpha(\alpha-1)}}
 \p^\mu\alpha\bigg] -2(\p_\mu\varphi_2)\p^\mu\varphi_2 \\
&\quad -\frac{1}{2}e^{-2\varphi_1} \big(e^{-2\varphi_2}F^1_{\mu\nu}F^{1\mu\nu}
 +e^{2\varphi_2}F^2_{\mu\nu}F^{2\mu\nu}\big)
 -e^{-2\varphi_1}\big(F^1_{\mu\nu}F^{2\mu\nu} +2F^4_{\mu\nu}F^{4\mu\nu}\big) \\
&\qquad +\frac{2\alpha-1}{\sqrt{\alpha(\alpha-1)}} e^{-2\varphi_1}\big(e^{-\varphi_2}F^1_{\mu\nu}
 +e^{\varphi_2}F^2_{\mu\nu}\big)F^{4\mu\nu} = 0 \, .
\end{split}\ee

These scalar field equations illustrate how the gauge fields act as effective potentials for the
scalar fields, creating a coupled system where the scalars and gauge fields mutually influence
each other's evolution. The complexity of these equations reflects the rich structure of the
$STU-W^2U$ model and underscores the challenges involved in constructing explicit exact solutions.

\subsection{Static non-extremal $STU-W^2U$ black hole}

We now present our main results: a class of new static non-extremal black hole solutions with four
independent electric charges in the $STU-W^2U$ model. The construction of this solution represents
a significant technical challenge, as no solution-generating technique is currently available for
this model. Our approach involves a combination of ans\"atz-based methods and direct brute-force
verification of the field equations.

\paragraph{Scalar field ans\"atz and metric structure:}
We begin with the following ans\"atz for the scalar fields, which generalizes the three-charge $STU$
case where $Z_4 = 0 = q_4$:
\be
\varphi_1 = \frac{1}{6} \ln \Big(\frac{Z_3^2}{Z_1Z_2 -Z_4^2}\Big) \, , \quad
\varphi_2 = \frac{1}{2} \ln \Big(\frac{Z_2}{Z_1}\Big) \, , \quad
\alpha = \frac{Z_1Z_2}{Z_1Z_2 -Z_4^2}\, ,
\ee
with the profile functions $Z_I$ given by:
\be
Z_i = 1 +\frac{q_i}{r^2}\, , \quad (i = 1, 2, 3)\, , \qquad Z_4 = \frac{q_4}{r^2} \, .
\ee
Here, the parameters $q_I$ are related to the physical charges of the solution. The appearance of
the combination $Z_1Z_2 -Z_4^2$ reflects the distinctive structure of the $STU-W^2U$ pre-potential.

For the metric, we employ a generalized ans\"atz that reduces to the known three-charge solution
when $Z_4 = q_4 = 0$:
\be
ds^2 = (Z_1Z_2 -Z_4^2)^{1/3}Z_3^{1/3} \bigg[ -\frac{f(r) dt^2}{(Z_1Z_2 -Z_4^2)Z_3}
 +\frac{dr^2}{f(r)} +r^2 d\Omega_3^2 \bigg] \, ,
\label{4csbh}
\ee
where $d\Omega_3^2$ is the metric on the unit 3-sphere as given by Eq. (\ref{3dsphere}). The
function $f(r)$ determines the horizon structure and asymptotic behavior of the solution.

\paragraph{Special cases and solution families:}
The remaining task is to specify the expressions for the four Abelian gauge fields, which divide
the solutions into different classes: super-symmetric BPS, extremal or non-extremal. We now present
the interested solutions in several special cases that illustrate the richness of the solution-space.

\subsubsection{Supersymmetric BPS case}

For the BPS case, which preserves some supersymmetry with $Q_I = \pi q_I/4$, we have the simplified
expression for the structure function: $f(r) = 1$, and the gauge filed 1-form potentials take the
particularly symmetric form:
\be
A_1 = \frac{\pm Z_2}{Z_1Z_2 -Z_4^2} dt \, , \quad A_2 = \frac{\pm Z_1}{Z_1Z_2 -Z_4^2} dt \, ,
\quad A_3 = \frac{\pm 1}{Z_3} dt \, , \quad A_4 = \frac{\pm Z_4}{Z_1Z_2 -Z_4^2} dt \, .
\ee

The function $f(r) = 1$ satisfies the differential equation:
\be
\frac{\p^2f(r)}{\p r^2} +\frac{7}{r}\frac{\p f(r)}{\p r} +\frac{8}{r^2}(f(r)-1) = 0 \, ,
\ee
which admits the more general solution $f(r) = 1 +f_2/r^2 +f_4/r^4$.

The above BPS solution resembles the four-charge static black ring solution \cite{CQG29-085006}
and represents the extremal limit where the horizon approach to the origin ($r = 0$).

\subsubsection{Special case: $q_2 = q_1$ and $p_2 = p_1$ ($Z_2 = Z_1$)}

When two of the charges are set to equal, the metric exhibit some enhanced symmetry. In this case,
the gauge potential 1-forms simplify to:
\be\begin{split}
A_1 &= A_2 = \frac{p_1Z_1 -p_4 Z_4}{r^2(Z_1^2 -Z_4^2)} dt \, , \\
A_3 &= \frac{p_3}{r^2Z_3} dt \, , \quad
A_4 = \frac{p_1Z_4 -p_4 Z_1}{r^2(Z_1^2 -Z_4^2)} dt \, ,
\end{split}\ee
and the metric function reads:
\bea
f(r) &=& 1 +\frac{2(q_1q_4 -p_1p_4)}{q_4r^2} +\Big(p_4^2 -q_4^2 +q_1^2 +p_1^2
 -\frac{2q_1p_1p_4}{q_4}\Big)\frac{1}{r^4} \nn \\
&=& 1 +\frac{2(q_1q_4 -p_1p_4)}{q_4r^2}Z_3 +\frac{p_3^2 -q_3^2}{r^4} \, ,
\eea
with one constraint condition controlling the constants $q_I$ and $p_I$:
\be
p_3^2 = p_4^2 -q_4^2 +(q_3 -q_1)^2 +p_1^2 +\frac{2(q_3 -q_1)p_1p_4}{q_4} \, .
\ee

\subsubsection{General case: $q_2 \neq q_1$ and $p_2 \neq p_1$ ($Z_2 \neq Z_1$)}

We now turn to the completely general and most interesting case with four independent electric
charges: $Q_I = \pi p_I/4$, and have:
\be\begin{split}
A_1 &= \frac{p_1Z_2 -p_4 Z_4}{r^2 (Z_1Z_2 -Z_4^2)}dt \, , \quad
A_2 = \frac{p_2 Z_1 -p_4 Z_4}{r^2 (Z_1Z_2 -Z_4^2)}dt \, , \\
A_3 &= \frac{p_3}{r^2 Z_3}dt \, , \quad
A_4 = \frac{q_4 (p_2 Z_1 -p_1Z_2)}{(q_1 -q_2)r^2 (Z_1Z_2 -Z_4^2)}dt \, ,
\end{split}\ee
and the metric function:
\bea
f(r) &=& 1 +\frac{q_1^2 -q_2^2 -p_1^2 +p_2^2}{(q_1 -q_2)r^2} +\Big(p_4^2 -q_4^2 +q_1q_2
 +\frac{q_1p_2^2 -q_2p_1^2}{q_1 -q_2}\Big)\frac{1}{r^4} \nn \\
&\equiv& 1 +\frac{q_1^2 -q_2^2 -p_1^2 +p_2^2}{(q_1 -q_2)r^2}Z_3 +\frac{p_3^2 -q_3^2}{r^4} \, ,
\label{ff0}
\eea
with the relation:
\be
p_4 = q_4 \frac{p_1 -p_2}{q_1 -q_2} \, ,
\ee
subject to a constraint condition among the eight constants $q_I$ and $p_I$:
\be
p_3^2 = p_4^2 -q_4^2 +(q_3 -q_2)(q_3 -q_1) +\frac{(q_3 -q_2)p_1^2 -(q_3 -q_1)p_2^2}{q_1 -q_2} \, .
\ee

A particularly suggestive choice to solve this constraint is given by:
\be\begin{split}
p_1 &= \sqrt{q_1^2 +2mq_1 +wq_4^2} \, , \quad
p_2 = \sqrt{q_2^2 +2mq_2 +wq_4^2} \, , \\
p_3 &= \sqrt{q_3^2 +2mq_3 +p_4^2 +(w-1)q_4^2} \, ,
\end{split}\ee
so the structure function is simplified to
\be
f(r) = 1 -\frac{2m}{r^2} +\frac{p_4^2 +(w-1)q_4^2}{r^4} \, ,
\label{ff1}
\ee
where $m$ is the mass parameter of the black hole, while $w$ is an arbitrary constant with two
particularly simple settings: $w=0$ or $w=1$. It should be pointed out that unlike
in the case of $STU$ model, the appearance of the $r^{-4}$ term in the function $f(r)$ (\ref{ff1})
indicates that no solution-generating technique is available for deriving the non-extremal static
four-charge solution from the five-dimensional Schwarzschild solution.

Clearly when $q_4 = p_4 = 0$, $q_i = 2ms_i^2$, $p_i = 2mc_is_i$, and $Z_i = h_i = 1 +2ms_i^2/r^2$,
($i=1,2,3$), our solution reduces to the static three-charge HMS black hole solution \cite{PLB383-151}
(after setting $r_0^2 = 2m$), demonstrating the consistency of our generalization.

\subsection{Thermodynamic properties of the general case}

The thermodynamic analysis of our four-charge static non-extremal black hole solution reveals a
richer structure that generalizes the well-known results for the three-charge HMS case. Obviously,
this solution is asymptotically flat since we have already chosen a clever gauge to let all the
three scalar fields vanish at the infinity.

Our black hole possesses a regular event horizon located at $r = r_+$, the largest root of $f(r_+)
= 0$. The Bekenstein-Hawking entropy, determined by the horizon area, is:
\be
S = \frac{1}{2}\pi^2 r_+^3\sqrt{\big(Z_1Z_2 -Z_4^2\big)Z_3}\Big|_{r = r_+} \, ,
\label{eq:entropy}
\ee
while the Hawking temperature is given by:
\be
T = \frac{\p_r f(r)}{2\big(Z_1Z_2 -Z_4^2\big)Z_3}\Big|_{r = r_+} \, .
\label{eq:temperature}
\ee

The electrostatic potentials computed at the horizon, conjugate to the electric charges, are:
\be\begin{split}
\Phi_1 &= \frac{p_1Z_2 -p_4Z_4}{r^2(Z_1Z_2 -Z_4^2)}\Big|_{r = r_+} \, , \quad
\Phi_2 = \frac{p_2 Z_1 -p_4 Z_4}{r^2(Z_1Z_2 -Z_4^2)}\Big|_{r = r_+} \, , \\
\Phi_3 &= \frac{p_3}{r^2Z_3}\Big|_{r = r_+} \, , \quad
\Phi_4 = \frac{q_4(p_2Z_1 -p_1Z_2)}{(q_1 -q_2)r^2(Z_1Z_2 -Z_4^2)}\Big|_{r = r_+} \, .
\end{split}\ee

The ADM mass and four electric charges are computed using the standard Komar and Gaussian integral
methods and are simply given by
\be
M = \frac{\pi}{4}(3m +q_1 +q_2 +q_3) \, , \quad
Q_I = \frac{\pi}{4} p_I \quad (I = 1,2,3,4) \, .
\ee

Remarkably, these thermodynamic quantities satisfy both the differential and integral forms of the
first law of black hole mechanics:
\be\begin{split}
d M &= TdS +\Phi_1dQ_1 +\Phi_2dQ_2 +\Phi_3dQ_3 -2\Phi_4dQ_4 \, , \\
M &= \frac{3}{2} TS +\Phi_1Q_1 +\Phi_2Q_2 +\Phi_3Q_3 -2\Phi_4Q_4 \, .
\end{split}\ee
The factor of `$-2$' in front of the terms: $\Phi_4 dQ_4$ and $\Phi_4Q_4$, is particularly noteworthy
and reflects the distinctive coupling of the fourth gauge field in the $STU$-$W^2U$ model, as evident
from the structure of the Chern-Simons term and the scalar kinetic couplings. The verification of
these relations provides a strong consistency check on our solution and demonstrates the internal
coherence of the thermodynamic description.

\section{Two extensions of the four-charge static non-extremal black hole solution}

Having established the general four-charge static non-extremal solution, we now present two natural
extensions that broaden the physical applicability of our results: the squashed (Klein-Kaluza)
version and an AdS$_5$ generalization.

\subsection{Squashing the horizons}

Black holes with squashed horizons have attracted considerable interest during the past years
due to their novel geometric properties. By directly employing the squashing transformation
\cite{PTP116-417} to our four-charge static black hole solution (\ref{4csbh}), we succeed in
extending it to include squashed horizon geometries. The metric for this generalization takes
the form:
\be\begin{split}
ds^2 &= \big(Z_1Z_2 -Z_4^2\big)^{1/3}Z_3^{1/3}\bigg[-\frac{f(r)}{(Z_1Z_2 -Z_4^2)Z_3} dt^2
 +\frac{k(r)^2}{f(r)} dr^2 \\
&\quad +\frac{k(r) r^2}{4}(d\vartheta^2 +\sin^2\vartheta\, d\hat\phi^2)
 +\frac{r^2}{4}\sigma_3^2 \bigg] \, ,
\end{split}\ee
where $\sigma_3 = d\hat\psi +\cos\vartheta\, d\hat\phi$, and the squashing function $k(r)$ is
given by:
\be
k(r) = \frac{r_\infty^4 -2mr_\infty^2 +p_4^2 +(w-1)q_4^2}{(r_\infty^2 -r^2)^2} \, .
\ee
The function $f(r)$ maintains its same form as given in Eq. (\ref{ff0}) or (\ref{ff1}). This
solution represents a Klein-Kaluza-type black hole with a horizon of a squashed 3-sphere,
that is, its spacetime is locally asymptotically flat and has a spatial infinity $R\times S^1
\hookrightarrow S^2$. However, because the scalar moduli does not vanishes asymptotically at
infinity, the first law should also include the contribution of the scalar hairs \cite{PRL77-4992}.
Rather, one can get a much simpler solution in which three scalar fields vanish asymptotically
at infinity, as did in refs. \cite{PLB726-404,GRG48-154}.

\subsection{Gauged supergravity extension}

The inclusion of a negative cosmological constant is of considerable physical interest,
particularly in the context of the AdS/CFT correspondence. In order to extend our solution
to the gauged supergravity theory, the following scalar potential must be added into the
Lagrangian:
\be
\cL_V = -2g_0^2\big[\sqrt{\alpha}(e^{-\varphi_2 -\varphi_1} +e^{\varphi_2 -\varphi_1})
 +e^{2\varphi_1}\big] = -2g_0^2 \big[(X^1 +X^2)X^3 +X^1X^2 -(X^4)^2\big] \, .
\ee
The static non-extremal AdS$_5$ black hole solution in this case is only modified via a
simple replacement:
\be
f(r) \to f(r) +g_0^2 r^2\big(Z_1Z_2 -Z_4^2\big)Z_3 \, .
\ee
This modification ensures that the solution asymptotically approaches AdS$_5$ spacetime with
a length scale $l = g_0^{-1}$. Just as the ungauged case, the above AdS$_5$ extension obviously
reduces the AdS$_5$ extension \cite{NPB553-317} of the HMS static $STU$ solution when the fourth
charge vanishes ($q_4 = 0$).

As far as its thermodynamics is concerned, many expressions retain a structure
similar to that of the ungauged case after suitable modifications. However, the precise form
of the first law of thermodynamics and the Smarr relation depends on the thermodynamic framework
adopted. If one adopts the formalism of the extended thermodynamic phase space, where the
cosmological constant is interpreted as a pressure, then the additional terms such as $VdP$
and $VP$ may contribute. A detailed analysis of its thermodynamic properties of the AdS version
within this framework has been delivered in a subsequent work \cite{2510.20164}. On the other
hand, if one defines the mass and electric charges as conserved quantities appropriately via
holographic renormalization (see Appendix A of ref. \cite{JHEP0523053}), then the first law
and Smarr relation may not contain any cosmological term. This is certainly what occurs in
the $STU$ model.\footnote{We thank our referee for kindly pointing out this
alternative perspective to us.}

These extensions demonstrate the robustness of our solution and its adaptability to different
physical contexts, opening up possibilities for further investigations in both asymptotically
flat and asymptotically AdS settings.

\section{Conclusions}

In this paper, we have studied a new model in the five-dimensional $\cN = 2$ supergravity theory
coupled to \emph{three} vector multiplets, which we refer to as the $STU-W^2U$ model. This model
generalizes the well-known $STU$ one and allows for the construction of a static black hole
solution with four independent electric charges. The solution reduces to the three-charge HMS
black hole when the fourth charge vanishes.

We have presented the full Lagrangian, field equations, and the explicit form of the static
non-extremal black hole solution. Its thermodynamic quantities---mass, entropy, temperature,
four electric charges, and their corresponding electrostatic potentials---have been computed
and are shown to satisfy both the differential and integral first laws of black hole thermodynamics.
We have also extended the solution to the case that includes squashed horizons or the case with
a nonzero negative cosmological constant.

Just as the case of the $STU$ model, the present $STU-W^2U$ model clearly admits various exact
solutions that generalize the non-extremal double-rotating (AdS$_5$) black hole \cite{PRD59-064005},
black ring \cite{hep-th/0612005}, and black lens \cite{PRD78-064062}, etc. A natural future direction
of the next step is to include two independent rotations to our static non-extremal black hole
solution, especially with two equal angular momenta for the relative easy case. The cases of two
different rotations that generalize the famous three-charge Cveti\v{c}-Youm solution \cite{NPB476-118}
and ``Wu black hole" \cite{PLB707-286} remain challenging and will be conducted in future work.
Perhaps the most challenging task is to pursue a 11-parameter solution that represents the
double-spinning non-extremal black ring with four independent electric charge and four different
dipole charge in the $STU-W^2U$ model, that extends the one found in ref. \cite{JHEP0814059} in
the five-dimensional $STU$ supergravity.

In addition to these directions, another interesting avenue for future research is to investigate the thermodynamic topological classification of black holes \cite{PRL129-191101,PRD107-024024,PRD107-084002,EPJC83-365,EPJC83-589,
PRD108-084041,JHEP0624213,PLB856-138919,CQG42-125007,PLB860-139163,PRD110-L081501,PRD111-L061501,
PRD112-124024} in the $STU-W^2U$ model. By analyzing the critical points, phase transitions, and associated topological charges of the black hole solutions, one may gain deeper insight into the global structure of the solution space and the relation between different black hole families. This could also provide a natural generalization of recent studies on thermodynamic topology in the $STU$ model \cite{JHEP0624213,PRD111-L061501} and may uncover universal properties that extend beyond specific solution ans\"atze.

\acknowledgments
We are grateful to the anonymous referee for useful suggestions. This work is
supported by the National Natural Science Foundation of China (NSFC) under Grant No. 12205243
and No. 12375053, and by the Sichuan Science and Technology Program under Grant No. 2026NSFSC0021.

\end{document}